\documentclass[12pt]{cernrep}
\usepackage{subfigure}
\usepackage{floatflt}
\usepackage{multirow}
\usepackage{amsmath}
\usepackage{amssymb}
\usepackage[english]{babel}
\usepackage{axodraw}
\usepackage{epsfig}
\usepackage{here}
\usepackage{booktabs}
\usepackage{cite}
\usepackage{here}
\usepackage{graphicx}
\usepackage{color}
\usepackage{slashed}
\newcommand{\SUSYBSMfile}[1]{#1}

\newcommand{\nonSUSYBSMfile}[1]{#1}

\newcommand{\ToolsBSMfile}[1]{#1}

\begin{document}

\setcounter{tocdepth}{0}
\pagestyle{myheadings}
\thispagestyle{empty}

\title{
LES HOUCHES ``PHYSICS AT TEV COLLIDERS 2005''
BEYOND THE STANDARD MODEL WORKING GROUP: SUMMARY REPORT
}

\author{
\underline{{\bf B.C.~Allanach}}$^{1}$,     
\underline{\bf C.~Grojean}$^{2,3}$,
\underline{\bf P.~Skands}$^{4}$,
			E.~Accomando$^{5}$,
			G.~Azuelos$^{6,7}$,
			H.~Baer$^{8}$,
			C.~Bal\'azs$^{9}$,
			G.~B\'elanger$^{10}$,
			K.~Benakli$^{11}$,
			F.~Boudjema$^{10}$
% ...
}

\institute{\vspace{1cm}
{\bf Editors} of proceedings in bold\\
\underline{convenor} of {\em Beyond the Standard Model} working group\\
$^1$ DAMTP, CMS, Wilberforce Road, Cambridge, CB3 0WA, UK\\ 
$^2$ SPhT, CEA-Saclay, Orme des Merisiers, F-91191 Gif-sur-Yvette Cedex, France\\ 
$^3$ Physics Department, CERN, CH-1211 Geneva 23, Switzerland\\ 
$^4$ Fermilab (FNAL), PO Box 500, Batavia, IL 60510, USA\\ 
$^{5}$ INFN, Sezione di Torino and Universit\`a di Torino, Dipartimento di Fisica Teorica, Italy\\ 
$^{6}$ Universit\'e de Montr\'eal, Canada\\ 
$^{7}$ TRIUMF, Vancouver, Canada\\ 
$^{8}$ Department of Physics, Florida State University, Tallahassee, FL 32306, USA\\ 
$^{9}$ HEP Division, Argonne National Laboratory, 9700 Cass Ave., Argonne, IL 60449, USA \\ 
$^{10}$ LAPTH, 9 Chemin de Bellevue, B.P. 110, Annecy-le-Vieux 74941, France\\ 
$^{11}$ LPTHE, Universit\'es de Paris VI et VII, France\\ 
$^{12}$ Moscow State University, Russia\\ 
$^{13}$ Institute for High Energy Phenomenology, Cornell University, Ithaca, NY 14853, USA\\ 
$^{14}$ Department of Physics and Astrophysics, University of Delhi, Delhi 110 007, India\\ 
$^{15}$ Universit\`a di Milano - Dipartimento di Fisica and Istituto Nazionale di Fisica Nucleare - Sezione di Milano, Via Celoria 16, I-20133 Milan, Italy\\ 
$^{16}$ Albert-Ludwigs Universit\"{a}t Freiburg, Physikalisches Institut, Hermann-Herder Str. 3, D-79104 Freiburg, Germany\\ 
$^{17}$ Universit\'e Cadi Ayyad, Facult\'e des Sciences Semlalia, B.P. 2390, Marrakech, Maroc\\ 
$^{18}$ LPT, Universit\'e de Paris XI, B\^at. 210, F-91405 Orsay Cedex, France\\ 
$^{19}$ Department of Physics, University of Oslo,Oslo, Norway\\ 
$^{20}$ Lancaster University, Lancaster LA1 4YB, UK \\ 
$^{21}$ Institut f\"ur Theoretische Physik und Astrophysik, Universit\"at   W\"urzburg, Germany\\ 
$^{22}$ Institute for Theoretical Physics, Univ. of Zurich, CH-8050 Zurich, Switzerland \\ 
$^{23}$ Indian Institute of Science, IISc, Bangalore, 560012, India\\ 
$^{24}$ LPC Clermont-Ferrand, Universit\'e Blaise Pascal, France\\ 
$^{25}$ Departament d'Estructura i Constituents de la Mat{\`e}ria, Facultat de F{\'\i}sica,  Universitat de Barcelona, Diagonal 647, E-08028 Barcelona, Catalonia, Spain\\ 
$^{26}$ Tata Institute of Fundamental Research, Homi Bhabha Road, Mumbai 400005, India\\ 
$^{27}$ MPI f\"ur Physik, Werner-Heisenberg-Institut, D--80805 M\"unchen, Germany\\ 
$^{28}$ Depto.\ de F\'isica Te\'orica, Universidad de Zaragoza, 50009 Zaragoza, Spain \\ 
$^{29}$ National Institute of Chemical Physics and Biophysics, Ravala 10, Tallinn 10144, Estonia\\ 
$^{30}$ High Energy Physics, Uppsala University, Box 535, S-751~21 Uppsala, Sweden\\ 
$^{31}$ LPTA, UMR5207-CNRS, Universit\'e Montpellier II, F-34095 Montpellier Cedex 5, France\\ 
$^{32}$ SLAC, Stanford University, Stanford, California 94409 USA\\ 
$^{33}$ Physics Div. 2, Institute of Particle and Nuclear Studies, KEK,  Tsukuba Japan\\ 
$^{34}$ Instytut Fizyki Teoretycznej, Uniwersytet Warszawski, PL-00681  Warsaw, Poland\\ 
$^{35}$ LAPP, 9 Chemin de Bellevue, B.P. 110, Annecy-le-Vieux 74941, France\\
$^{36}$ Brown University, Providence, Rhode Island, USA\\
$^{37}$ CTP, School of Physics, Seoul National University, Seoul 151-747, Korea\\ 
$^{38}$ Physics Department, Mount Allison University,  Sackville NB, E4L 1E6 Canada\\ 
$^{39}$ Enrico Fermi Institute, University of Chicago, 5640 S. Ellis Ave., Chicago, IL 60637, USA \\ 
$^{40}$ Department of Physics and Astronomy, University of Glasgow, Glasgow G12 8QQ, UK\\ 
$^{41}$ IPN Lyon, 69622 Villeurbanne, France\\ 
$^{42}$ School of Physics and Astronomy, University of Southampton, SO17 1BJ, UK\\ 
$^{43}$ Particle Physics Division, Rutherford Appleton Laboratory, Oxon OX11 0QX, UK \\ 
$^{44}$ Department of Physics, University of Michigan, Ann Arbor, MI 48109, USA \\ 
$^{45}$ LAL, Universit\'e de Paris-Sud, Orsay Cedex, France\\ 
$^{46}$ Deutsches Elektronen-Synchrotron DESY, D--15738 Zeuthen, Germany\\ 
$^{47}$ SPP, DAPNIA, CEA-Saclay, F-91191 Gif-sur-Yvette Cedex, France\\ %
$^{48}$ School of Physics and Astronomy, University of Manchester, Manchester M13  9PL, UK\\
$^{49}$ University of Edinburgh, GB\\ 
$^{50}$ INFN, Sezione di Pavia, Via Bassi 6, I-27100 Pavia, Italy \\ 
$^{51}$ University of Oklahoma, USA\\ 
$^{52}$ Instituto de F\'{\i}sica Corpuscular, C.S.I.C., Val\`encia, Spain\\ 
$^{53}$ Skobeltsyn Inst. of Nuclear Physics, Moscow State Univ., Moscow 119992, Russia\\ 
$^{54}$ Dept. of Physics and Astronomy, University of Rochester, NY, USA\\
$^{55}$ Dept.\ of Physics and Technology, University of Bergen, N-5007 Bergen, Norway \\ 
$^{56}$ DESY Theory Group, Notkestr. 85, D-22603 Hamburg, Germany\\
$^{57}$ IPPP, University of Durham, Durham DH1~3LE, UK\\ 
$^{58}$ Physical Research Laboratory, Ahmedabad, India\\ 
$^{59}$ Paul Scherrer Institut, CH--5232 Villigen PSI, Switzerland\\ 
$^{60}$ Zweites Physikalisches Institut der Universit\"at, D-37077 G\"ottingen, Germany\\ 
$^{61}$ Institute for Theoretical Physics, TU Dresden, 01062, Germany\\ 
$^{62}$ Joint Institute for Nuclear Research (JINR), 143980, Dubna, Russia\\ 
$^{63}$ INFN, Sezione di Roma and Universit\`a di Roma ``La Sapienza", I-00185 Rome, Italy\\ 
$^{64}$ University of California, Berkeley, USA \\ 
$^{65}$ Department of Physics, Okayama University, Okayama, 700-8530, Japan\\
$^{66}$ IEKP, Universit\"at Karlsruhe (TH), P.O. Box 6980, 76128 Karlsruhe, Germany\\ 
$^{67}$ SINP, Lomonosov Moscow State University, 119992 Moscow , Russia\\ 
\vspace{1cm}
}

\maketitle

\begin{abstract}
The work contained herein constitutes a 
report of the ``Beyond the Standard Model'' working group for the Workshop 
``Physics at TeV Colliders", Les Houches, France, 2--20 May, 2005.
We present reviews of current topics as well as original research carried out
for the workshop. Supersymmetric and non-supersymmetric models are studied, as
well as computational tools designed in order to facilitate their
phenomenology.
\end{abstract}

\vfill

\section*{Acknowledgements}

We would like to heartily thank the funding bodies, organisers, staff and other
participants of the Les Houches workshop for providing a stimulating and
lively environment in which to work. 

\newpage

\tableofcontents

\newpage

\newcommand{\superpart}[1]{\clearpage \thispagestyle{empty}  \vphantom{blabla}\vspace{5cm} \centerline{\huge #1} \clearpage
 \addcontentsline{toc}{part}{\protect\numberline{} \hspace{-2cm}#1}%
}

\renewcommand{\thepart}{\arabic{part}}
\renewcommand{\thesubsubsection}{\arabic{section}.\arabic{subsection}.\arabic{subsubsection}}

\superpart{BSM SUSY}

\input \SUSYBSMfile{bsmSusyIntro.tex}
\setcounter{figure}{0}
\setcounter{table}{0}
\setcounter{section}{0}
\setcounter{equation}{0}
\clearpage

\input \SUSYBSMfile{focusATLAS.tex}
\setcounter{figure}{0}
\setcounter{table}{0}
\setcounter{section}{0}
\setcounter{equation}{0}
\clearpage

\input \SUSYBSMfile{focus.tex}
\setcounter{figure}{0}
\setcounter{table}{0}
\setcounter{section}{0}
\setcounter{equation}{0}
\clearpage

\input \SUSYBSMfile{nazila.tex}
\setcounter{figure}{0}
\setcounter{table}{0}
\setcounter{section}{0}
\setcounter{equation}{0}
\clearpage

\input \SUSYBSMfile{trilepton.tex}
\setcounter{figure}{0}
\setcounter{table}{0}
\setcounter{section}{0}
\setcounter{equation}{0}
\clearpage

\input \SUSYBSMfile{zhukov.tex}
\setcounter{figure}{0}
\setcounter{table}{0}
\setcounter{section}{0}
\setcounter{equation}{0}
\clearpage

\input \SUSYBSMfile{cp.tex}
\setcounter{figure}{0}
\setcounter{table}{0}
\setcounter{section}{0}
\setcounter{equation}{0}
\clearpage

\input \SUSYBSMfile{lightstop.tex}
\setcounter{figure}{0}
\setcounter{table}{0}
\setcounter{section}{0}
\setcounter{equation}{0}
\clearpage

\input \SUSYBSMfile{nmssm.tex}
\setcounter{figure}{0}
\setcounter{table}{0}
\setcounter{section}{0}
\setcounter{equation}{0}
\clearpage

\input \SUSYBSMfile{split.tex}
\setcounter{figure}{0}
\setcounter{table}{0}
\setcounter{section}{0}
\setcounter{equation}{0}
\clearpage

\input \SUSYBSMfile{split_Benakli.tex}
\setcounter{figure}{0}
\setcounter{table}{0}
\setcounter{section}{0}
\setcounter{equation}{0}
\clearpage

\input \SUSYBSMfile{gluino_cascade.tex}
\setcounter{figure}{0}
\setcounter{table}{0}
\setcounter{section}{0}
\setcounter{equation}{0}
\clearpage

\input \SUSYBSMfile{cphiggs.tex}
\setcounter{figure}{0}
\setcounter{table}{0}
\setcounter{section}{0}
\setcounter{equation}{0}
\clearpage

\input \SUSYBSMfile{m0.tex}
\setcounter{figure}{0}
\setcounter{table}{0}
\setcounter{section}{0}
\setcounter{equation}{0}
\clearpage

\superpart{TOOLS}

\input \ToolsBSMfile{repository.tex}
\setcounter{figure}{0}
\setcounter{table}{0}
\setcounter{section}{0}
\setcounter{equation}{0}
\clearpage

\input \ToolsBSMfile{slha2.tex}
\setcounter{figure}{0}
\setcounter{table}{0}
\setcounter{section}{0}
\setcounter{equation}{0}
\clearpage

\input \ToolsBSMfile{pythia-ued.tex}
\setcounter{figure}{0}
\setcounter{table}{0}
\setcounter{section}{0}
\setcounter{equation}{0}
\clearpage

\input \ToolsBSMfile{nmssmgen.tex}
\setcounter{figure}{0}
\setcounter{table}{0}
\setcounter{section}{0}
\setcounter{equation}{0}
\clearpage

\input \ToolsBSMfile{sherpa.tex}
\setcounter{figure}{0}
\setcounter{table}{0}
\setcounter{section}{0}
\setcounter{equation}{0}
\clearpage

\input \ToolsBSMfile{FH.tex}
\setcounter{figure}{0}
\setcounter{table}{0}
\setcounter{section}{0}
\setcounter{equation}{0}
\clearpage

\input \ToolsBSMfile{micrOMEGAs.tex}
\setcounter{figure}{0}
\setcounter{table}{0}
\setcounter{section}{0}
\setcounter{equation}{0}
\clearpage

\superpart{NON-SUSY BSM}

\input \nonSUSYBSMfile{bsmNonSusyIntro.tex}
\setcounter{figure}{0}
\setcounter{table}{0}
\setcounter{section}{0}
\setcounter{equation}{0}
\clearpage

\input \nonSUSYBSMfile{UED-II.tex}
\setcounter{figure}{0}
\setcounter{table}{0}
\setcounter{section}{0}
\setcounter{equation}{0}
\clearpage

\input \nonSUSYBSMfile{UED.tex}
\setcounter{figure}{0}
\setcounter{table}{0}
\setcounter{section}{0}
\setcounter{subsubsection}{0}
\setcounter{equation}{0}
\clearpage

\input \nonSUSYBSMfile{KKDM.tex}
\setcounter{figure}{0}
\setcounter{table}{0}
\setcounter{section}{0}
\setcounter{equation}{0}
\clearpage

\input \nonSUSYBSMfile{ghu.tex}
\setcounter{figure}{0}
\setcounter{table}{0}
\setcounter{section}{0}
\setcounter{equation}{0}
\clearpage

\input \nonSUSYBSMfile{LH.tex}
\setcounter{figure}{0}
\setcounter{table}{0}
\setcounter{section}{0}
\setcounter{equation}{0}
\clearpage

\input \nonSUSYBSMfile{higgspp.tex}
\setcounter{figure}{0}
\setcounter{table}{0}
\setcounter{section}{0}
\setcounter{equation}{0}
\clearpage

\input \nonSUSYBSMfile{topol.tex}
\setcounter{figure}{0}
\setcounter{table}{0}
\setcounter{section}{0}
\setcounter{equation}{0}
\clearpage

\input \nonSUSYBSMfile{tbh.tex}
\setcounter{figure}{0}
\setcounter{table}{0}
\setcounter{section}{0}
\setcounter{equation}{0}
\clearpage

\input \nonSUSYBSMfile{diphoton.tex}
\setcounter{figure}{0}
\setcounter{table}{0}
\setcounter{section}{0}
\setcounter{equation}{0}
\clearpage

\input \nonSUSYBSMfile{hless.tex}
\setcounter{figure}{0}
\setcounter{table}{0}
\setcounter{section}{0}
\setcounter{equation}{0}
\clearpage

\input \nonSUSYBSMfile{WZ.tex}
\setcounter{figure}{0}
\setcounter{table}{0}
\setcounter{section}{0}
\setcounter{equation}{0}
\clearpage

\bibliography{bsmreport}
\end{document}